\documentclass[12pt]{article}

\usepackage{siunitx}
\usepackage{graphicx}
\usepackage{float}
\usepackage{caption}
\usepackage{cite}

\makeatletter
\newcommand{\address}[1]{%
  \g@addto@macro\@author{\\[0.4em]\normalsize\itshape #1}%
}
\newcommand{\ead}[1]{%
  \g@addto@macro\@author{\\[0.4em]\normalsize\ttfamily #1}%
}
\makeatother

\title{\bfseries Subwavelength Phase Engineering Deep Inside Silicon}

\author{%
  \parbox{\dimexpr\textwidth-2\tabcolsep\relax}{\centering        
    Mehmet Bütün$^{1}$, Alperen Saltik$^{1}$ and Onur Tokel$^{1,2}$\\[0.8em]
    \textit{$^{1}$ Department of Physics, Bilkent University, Ankara, Turkey}\\
    \textit{$^{2}$ UNAM – National Nanotechnology Research Center and Institute of Materials Science and Nanotechnology,\\%
    Bilkent University, Ankara, Turkey}\\[0.6em]
    \texttt{otokel@bilkent.edu.tr}%
  }%
}

\date{}                 
\begin{document}
\maketitle

\begin{abstract}
Recent advances in three-dimensional laser writing have enabled direct nanostructuring deep within silicon, unlocking a volumetric design space previously inaccessible to surface-bound nanophotonic devices. Here, we introduce subwavelength phase engineering inside crystalline silicon, offering a novel strategy for integrated photonics. We design and numerically demonstrate a volumetric metaoptic monolithically embedded within the bulk, achieving full 2$\pi$ phase control at telecommunication wavelengths, with simulated transmission efficiencies reaching \SI{90}{\percent}. The architecture is guided by a semi-analytical Fabry–P\'erot model and validated through full-wave simulations. Arrays of 250-nm‑wide metaatoms spaced at 300–410 nm pitch yield a focusing efficiency of \SI{70}{\percent}. With the wafer surface left pristine, this platform can potentially enable co-integration with electronics, MEMS/NEMS, and conventional metasurfaces. Moreover, the method is directly transferable to other transparent dielectrics compatible with ultrafast laser writing. These results establish a CMOS-compatible blueprint for three-dimensional nanophotonics and multi-level integration within the wafer.
\end{abstract}


\maketitle

\section{Introduction} 
Silicon (Si) is a foundational material in electronics, photovoltaics and Si-photonics, where planar nanolithography is routinely employed to sculpt state-of-the-art nanophotonic devices on the wafer surface \cite{Staude:2017}. Exploiting the bulk of Si, rather than its surface, would unlock a fundamentally different platform for optical control \cite{Chambonneau2021_review}. Three-dimensional (3D)  nonlinear laser lithography offers a direct and mask-free approach to functionalize the wafer interior \cite{tokel2017chip,Ahsan,saltik2023laser}. These in-chip or in-volume techniques harness nonlinear absorption of laser pulses and their feedback-driven interactions to induce structural modifications deep within the crystal \cite{tokel2017chip,Ahsan,saltik2023laser}, enabling precise volumetric patterning \cite{tokel2017chip,Rodenas2019,Xu:2022he,Supstealth}. Importantly, this process preserves the wafer surface, allowing compatibility with existing device integration strategies \cite{tokel2017chip}. 

The laser-written regions inside Si have already enabled a range of volumetric photonic components, including waveguides, lenses, gratings, waveplates, and both Fourier and Fresnel holograms along with applications in optical data storage \cite{tokel2017chip,saltik2023laser,turnali2019laser,picosecond_nolte,Chanal2017,chambonneau2018inscribing, wang2021curved,alberucci2020depth,Bütan, Tokel2014_info_encod}. The emerging depth dimension has also been shown to boost the efficiency of micro-optics. For instance, the efficiency of diffraction gratings is significantly improved with effective field enhancement achieved over a 3D optical lattice formed inside the wafer \cite{Bütan}. Such an improvement would not be possible with single-level fabrication \cite{Bütan}. While these exciting advances have focused on micro-structuring, the recent discovery of volumetric nano-fabrication in silicon is envisioned to enable nanophotonics or metaoptics capabilities directly inside the wafer \cite{sabet2023laser}. In order to realize this potential and guide the next generation of experiments, it is essential to develop optical architectures that are inherently compatible with the depth degree-of-freedom. 

Metamaterials, composed of subwavelength building blocks known as metaatoms, \cite{chen2016review, Roadmap_for_Optical_Metasurfaces} would be appropriately challenging to create with in-chip approaches. The metaatoms are engineered to manipulate electromagnetic waves for advanced functionality \cite{shelby2001experimental}. Due to the numerous challenges in 3D fabrication, the field focuses on metasurfaces. These are more practical, as they require two-dimensional patterning with subwavelength resolution and pitch \cite{3D_Printed_Inverse_Designed_Metaoptics,li2018metasurfaces}. In contrast, 3D laser nano-fabrication is uniquely suitable to exploit the bulk \cite{sabet2023laser}. 

Although early developments in metaoptics focused on metallic resonators \cite{quevedoteruel_2019_roadmap}, recent efforts have shifted toward dielectric materials such as silicon, which offer high refractive index contrasts and low optical losses \cite{jahani_2016_alldielectric,mm_metaoptics,Adi2024,yu_2011_light}. These devices have been adopted for numerous functionalities, with control over the phase, amplitude, and polarization of light \cite{capasso_holey_lens,nature_review,genevet_2017_recent}. Common metasurface implementations include arrays of nanopillars patterned on planar substrates \cite{nature_review,Ji_Li_Wang_Li,Zhou_Fan}, or alternatively, subwavelength holes with varying diameters etched into a host material \cite{capasso_holey_lens, ishii_2012_holeymetal}. The optical response is governed by the height, pitch, and geometry of these nanostructures, facilitating enhanced light–matter interaction and advanced optics such as flat lenses  \cite{genevet_2017_recent,ishii_2012_holeymetal}. Despite significant progress, such metaoptics remain confined to surface-bound architectures, with their practical deployment hindered by complex nanofabrication demands, structural fragility, and susceptibility to environmental exposure \cite{yu2014flat}.

In this work, we introduce a design paradigm that leverages the volumetric nature of laser writing \cite{sabet2023laser} to enable monolithic, buried metaoptics compatible with conventional Si-photonics and metasurface platforms. By relocating optical functionality from the surface to the interior, this strategy mitigates contamination and packaging constraints while introducing a previously inaccessible integration dimension. Specifically, we present a framework for subwavelength phase engineering deep inside silicon. Continuous tiling of resonant metaatoms enables the construction of a hyperbolic-phase metalens with an aperture of 100 $\mu$m and a focal length of 120 $\mu$m. Simulations predict full 2$\pi$ phase coverage, up to 90\% transmission for individual metaatoms, and a focusing efficiency of 70\%  at the wavelength of 1.55 $\mu$m. 

These results establish a foundation for extending metaoptics into the bulk of crystalline silicon and, more broadly, for enabling multi-level three-dimensional photonic integration in semiconductor platforms through in-chip fabrication.

\begin{figure}[H]
\centering
\includegraphics[scale=0.4]{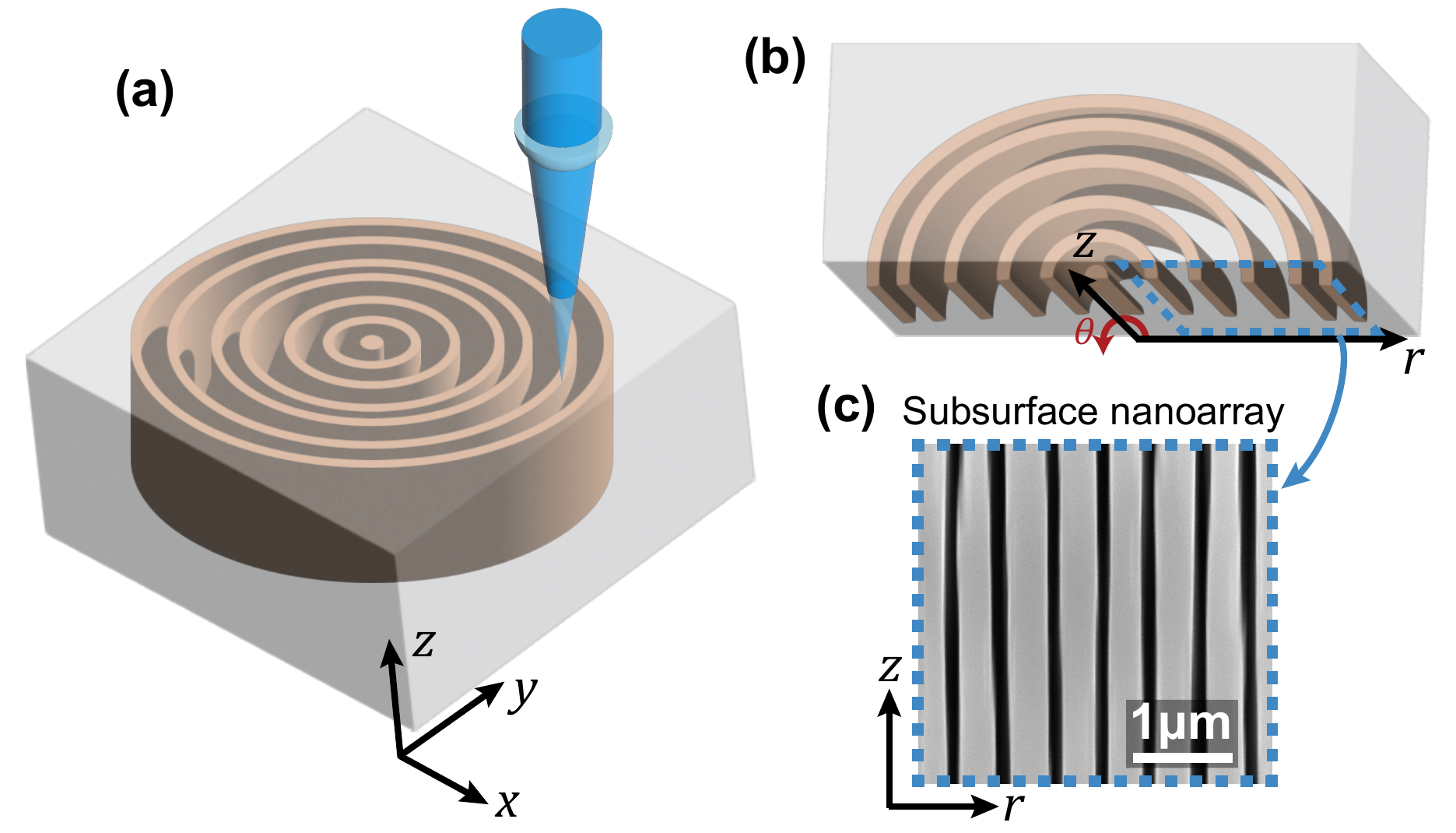}
\caption{\textrm{\textbf{The concept of in-chip metaoptics.} (\textbf{a}) Using three-dimensional laser writing, subsurface patterns with subwavelength features are created deep inside Si. Metaatoms are designed to create a particular phase pattern, enabling plug-and-play functionality. (\textbf{b}) A representative metalens inside Si. These may also be integrated with conventional metasurfaces. (\textbf{c}) Scanning electron microscopy image of a cross-section of the direct-laser-written nano-structures buried inside silicon. The sample is etched briefly, revealing a regular array with a feature size of $154~\pm~20$ nm and a period of 611 \(\pm\) 47 nm. A ns-pulsed laser of 1.55-\(\mu\)m wavelength, 5-\(\mu\)J pulse energy, linear polarization, and Bessel-profile is used.}}
\label{fig:1}
\end{figure} 

We build on the emerging foundation of in-chip laser nanostructuring, aiming for a design compatible with current subsurface fabrication capabilities (Fig. \ref{fig:1}). \cite{sabet2023laser} State-of-the-art laser writing in silicon has achieved resolutions down to 100 nm and pitch values as low as 300 nm \cite{sabet2023laser}. We show that optical functionality emerges from light interacting with such laser-written nanostructures, with phase shifts computed using finite-difference time-domain (FDTD) simulations across a range of pitch and height values. Our target is full 2$\pi$ phase modulation for advanced wavefront control. Based on these simulations, we design and numerically evaluate a subsurface metalens (Figs. \ref{fig:1}a-\ref{fig:1}b).  The required feature sizes and periodic nanopatterning fall within the capabilities of current experimental techniques across various contexts. As a proof of principle, we fabricate buried nanoarrays inside Si with laser pulses of \(\varepsilon_p\) = $5 ~\text{\textmu J}$, linear polarization, and Bessel-type beam profile (Fig. \ref{fig:1}c). The fabrication and characterization procedures of nanopatterns are detailed in ref. \cite{sabet2023laser}. Scanning electron microscope (SEM) analysis reveals periodic buried structures with feature size of $154~\pm~20$ nm and a pitch of $611~\pm~47$ nm (Fig. \ref{fig:1}c). These results highlight the feasibility of high-aspect-ratio periodic nanostructuring in silicon. Our simulations further demonstrate their optical potential in 3D nano-engineering. We conclude with a discussion of current limitations and future directions toward the realization of 3D Si-nanophotonics. 


\section{Design of Subsurface Metaatoms}

\begin{figure}[H]
\centering
\includegraphics[scale=0.42]{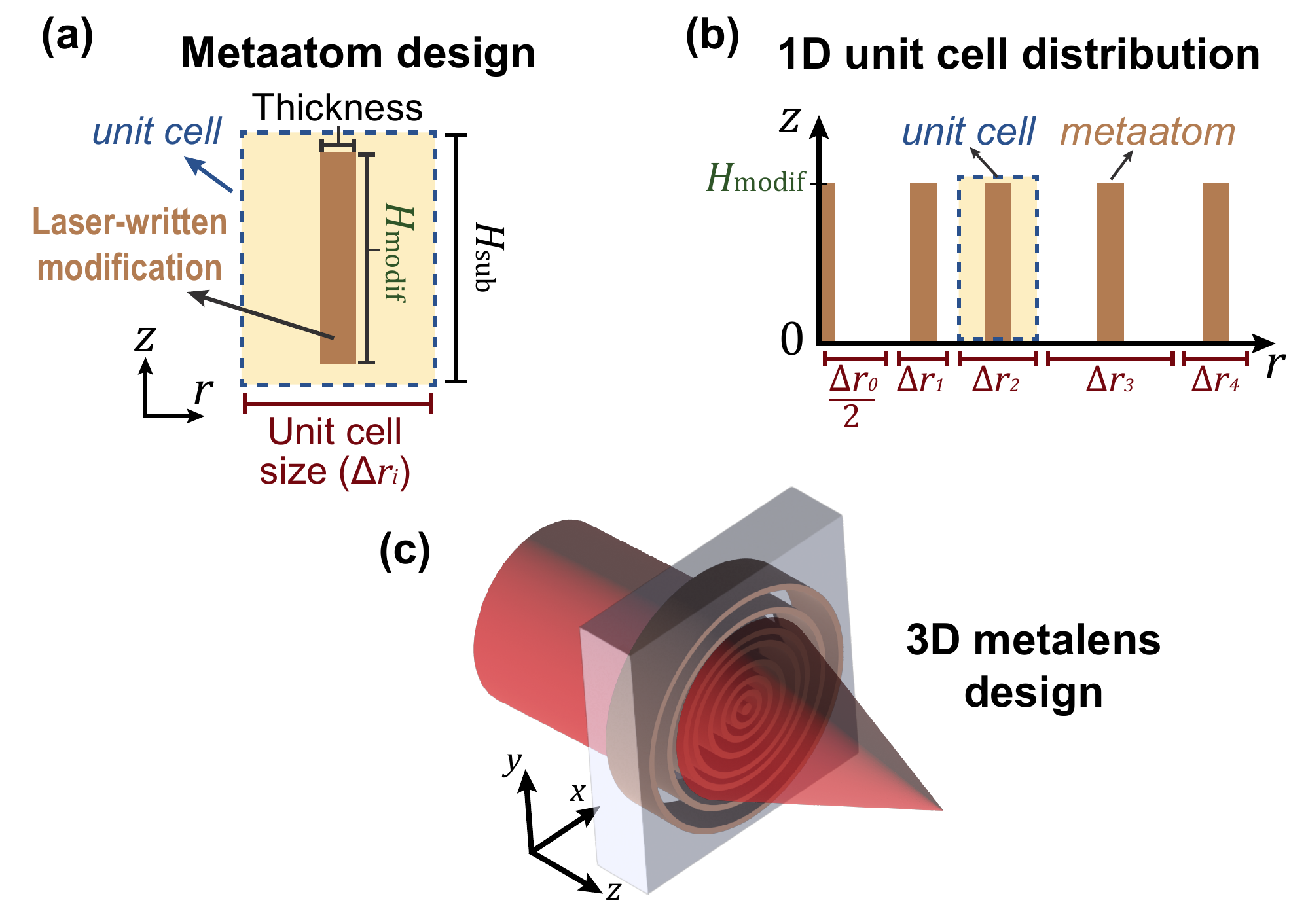}
\caption{\textrm{\textbf{Metaoptics buried inside Si.} (\textbf{a}) The metaatoms and unit cells from the crosssection of 3D optics. Metaatom width is ($\delta l$) assumed to be small enough to consider them as rectangular prisms with thickness of $\zeta$ = 250 nm and height $H_\text{modif}$. (\textbf{b}) Metaatom distribution over the $r$ coordinate. The unit cell ($\Delta r_i$) and patterning over $r$ are selected based on the targeted phase map of the lens. (\textbf{c}) 3D metalens is formed by continuous tiling of the pattern in (b) over the azimuthal angle ($\theta$), resulting in slightly elliptical forms.}}
\label{fig:2}
\end{figure}

Metasurfaces consist of basic units, $\textit{i.e.}$, metaatoms, engineered to impart complex optical functionality \cite{nature_review, genevet_2017_recent}. Here we pursue a distinct paradigm in which the metaatoms are buried within crystalline silicon. In contrast to conventional surface-bound fabrication schemes \cite{capasso_holey_lens}, our approach relies on direct laser writing \cite{sabet2023laser}. A novel seeding-based mechanism enables sub-wavelength features with aspect ratios ($>$1000) and allows multilevel nanostructuring deep inside silicon \cite{sabet2023laser}. Accordingly, each metaatom is modeled as a laser-written structure of thickness $\zeta$ and height $H_{\text{modif}}$ embedded in the crystal (Fig. \ref{fig:2}a). The lateral position, depth, feature size, and unit-cell geometry of the patterns can be controlled using holographic structuring and beam positioning techniques \cite{sabet2023laser}. This freedom already allows diverse design options, \textit{e.g.}, high-efficiency volume nano-gratings  \cite{sabet2023laser}.

The unit cells are to be tiled over a volume resulting in the novel architecture (Figs. \ref{fig:2}b-\ref{fig:2}c). The tiling pattern and feature dimensions are selected according to the target wavelength and desired optical functionality. We adopt the pitch as the control parameter for phase, as it can be adjusted in experiments. In order to create the circular symmetry of a lens, we exploit continuous laser-written arrays. This is a unique feature of laser-writing inside Si \cite{sabet2023laser}. In our preliminary experiments, we find that it is possible to fabricate continuous as well as curved nano-structures inside Si. The resulting pattern (Fig. \ref{fig:2}c) resembles a Fresnel lens, however, the operating principles are distinct. The former is based on Fourier optics, whereas our device exploits resonant coupling of wavelength-scale metaatoms \cite{Gao2019}.

The optical index landscape for laser-modified silicon is evolving rapidly, and remains a challenge to pin down quantitatively. Currently the index contrast is modest at the nanoscale ($|\Delta n| \approx 10^{-2}$) \cite{sabet2023laser}. Closer structural analyses, however, reveal two distinct microphases, localized amorphous pockets and nanovoids, that substantially broaden the attainable index range \cite{Chambonneau2021_review,sabet2023laser,amorphous_si}. The amorphous phase supports $\Delta n$ $\approx$ +0.25, while void formation can yield $\Delta n$ $\approx$ -2.5 \cite{sabet2023laser,de2002amorphous}. Indeed, ultrafast-pulse studies in Si already achieved partial amorphization with $\Delta n$ $\approx$ +0.1 \cite{Grojo2024PlasmaSeeds}. Moreover, tailoring the pulse width permits access to both positive  \cite{amorphous_si,Femtosilicon2017,nolte_waveguide}, and negative-contrast regimes\cite{Chambonneau2021_review,Chambonneau16}. A complementary route exploits the discovery of stress-induced birefringence to fabricate the first waveplates embedded in Si\cite{saltik2023laser}. Taken together, these results indicate that laser-based refractive-index engineering offers a path forward, and we presume controlled amorphization may be achieved with refined techniques \cite{tokel2017chip,amorphous_si,Grojo2024PlasmaSeeds}. Thus, we first investigated the case of $\Delta n$ = 0.25, followed by $|\Delta n|$ = $10^{-2}$. In both cases, we find full 2$\pi$ phase modulation may be possible, suggesting advanced photonics capabilities through judicious combination of existing features, emerging architectures and refractive index engineering.

\section{Theory and Simulations}
We employ the finite-difference time-domain (FDTD) method to model both the individual metaatoms and the complete metalens. The optical response of a single metaatom is first obtained with a 2D FDTD solver. In this configuration a linearly-polarized plane wave propagates along the \textit{z}-axis (Fig. \ref{fig:2}a). Periodic boundary conditions (PBCs) are imposed on the $r_{\min}$ and $r_{\max}$ boundaries, and perfectly matched layers (PMLs) are applied at $z_{\min}$, $z_{\max}$ to absorb outgoing waves. The resulting transmission and phase-shift spectra provide the lookup table used in the metalens design. The full device is then analyzed with a 3-D FDTD simulation that employs the same PML treatment on all outer boundaries.

\begin{table}[t]
\centering
\captionsetup{labelformat=empty} 
\caption{ \textbf{Table 1:} Metaatom design parameters.}
\begin{tabular}{cc}
\hline
\small \rmfamily Properties \par & \small \rmfamily Values \par \\
\hline
\small \rmfamily Material \par  &\small \rmfamily Silicon \par\\
\small \rmfamily Wavelength \par & \small \rmfamily $\lambda$ = 1550 \text nm\par\\
\small \rmfamily Wavelength in medium \par & \small \rmfamily $\lambda_{\text{Si}}$ = 445 \text nm\par\\
\small \rmfamily Modification thickness \par &\small \rmfamily  $\zeta$ = 250 nm  \par\\
\small \rmfamily Substrate height \par &\small \rmfamily $H_{\text{sub}}$ = 80 $\times \,\lambda_{\text{Si}}$ (35.63 \textmu m)  \par\\
\small \rmfamily Modification height \par &\small \rmfamily $H_{\text{modif}}$ = 60 $\times \,\lambda_{\text{Si}}$ (26.72 \textmu m)  \par\\
\small \rmfamily Refractive index contrast\par &\small \rmfamily $\Delta n$ = 0.25 \par\\
\hline
\end{tabular}
  \label{table 1}
\end{table}

\begin{figure}[H]
\centering
\includegraphics[width=0.92\linewidth]{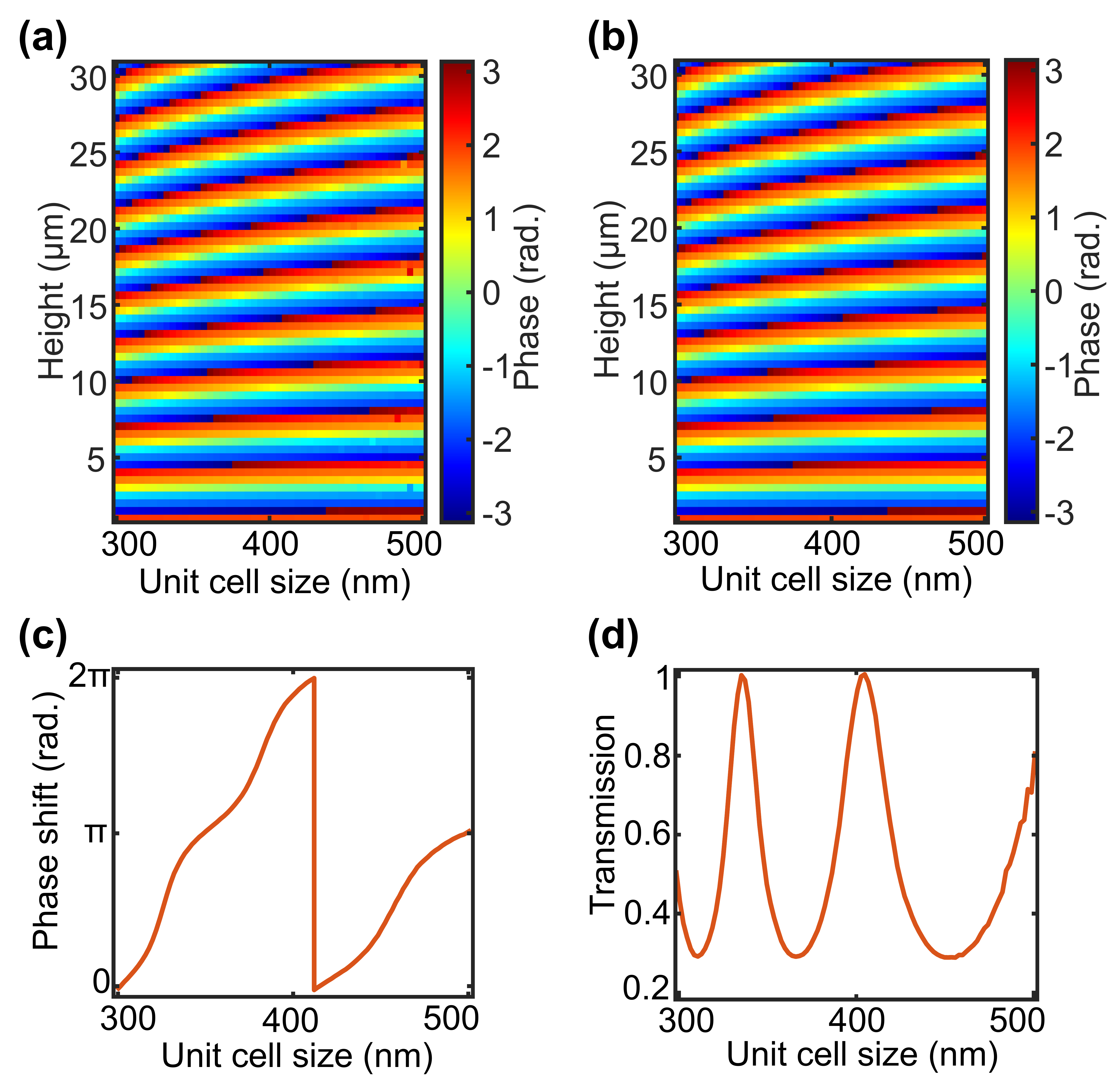}
\caption{\textrm{\textbf{Numerical and semi-analytical analyses of metaatoms for \boldmath$\Delta n$ = 0.25.} (\textbf{a}) FDTD simulation of phase shift for 300 to 500 nm unit cell size and 0 to 30 \textmu m heights. (\textbf{b}) Semi-analytical estimate from Eq. \ref{eq:fp}. We assume $60 \times \lambda_{\text{Si}}$ for the modification and substrate height in (a) and (b). (\textbf{c}) FDTD simulation of phase shift for $H_\text{modif}$ = $60 \times \lambda_{\text{Si}}$ and $H_\text{sub}$ = $80 \times \lambda_{\text{Si}}$ fully buried subsurface metaatoms. (\textbf{d}) FDTD simulation of transmission under normal incidence.}}
\label{fig:3}
\end{figure}

The parameters for the first metaatom design are summarized in Table \ref{table 1}. All simulations target the near-infrared wavelength $\lambda$ = 1.55 \textmu m, well inside the transparency window of Si ($>1 ~\text{\textmu m}$). Throughout the simulations, the metaatom height ($H_\text{modif} = 60~\lambda_{\text{Si}}$) and thickness ($\zeta$ = 250 nm) are held constant, while the phase modulation is recorded as a function of the unit cell size ($\Delta r_i$). We sweep $\Delta r_i$ in the range of 300 nm - 500 nm, a range that supports strong resonant coupling, while suppressing higher diffraction orders (Fig. \ref{fig:3}) \cite{Engelberg2020}. FDTD simulations with \textit{x}-polarized light yield the phase modulation map given in (Fig. \ref{fig:3}a), indicating smooth phase control across the selected pitch range. We observe full 2$\pi$ phase shift modulation (Fig. \ref{fig:3}c), with the transmission profile given in (Fig. \ref{fig:3}d).

As a secondary approach to ascertain the phase modulation, we developed a semi-analytical model tailored to our geometry. When neighbouring metaatoms are closely spaced, the ensemble can be approximated as a homogeneous slab of effective refractive index $n_{\textrm{eff}}$. In that limit, the phase and amplitude response are governed by the Fabry–P\'erot transmission equation (Eq. \ref{eq:fp}) \cite{capasso_holey_lens}. The complex transmission amplitude ($t_{\text{FP}}$) is then given as, \begin{equation}
t_{\text{FP}}=\frac{4\left(n_{\text{eff}} / n_{0}\right) e^{in_{\text{eff}}k_{0} H}}{\left(n_{\text{eff}} / n_{0}+1\right)^{2}-\left(n_{\text{eff}} / n_{0}-1\right)^{2} e^{2 i n_{\text{eff}}k_{0} H}} \, ,
\label{eq:fp}
\end{equation}
where $n_{\text{eff}}$ is the effective refractive index of the dominant mode, $n_0$ is the refractive index of background, $k_0$ is the wavenumber, and $H$ is the height of the resonator. The phase shift would correspond to $\arg$($t_{\textrm{FP}}$), and the transmission to $|t_{\text{FP}}|$. The effective refractive index ($n_{\text{eff}}$) is extracted from a mode analysis performed with the Lumerical MODE Solver (Supplementary Fig. S1). The predicted phase map (Fig. \ref{fig:3}b) matches the numerical results (Fig. \ref{fig:3}a) to within 0.1 rad (Supplementary Figs. S2-S3) confirming that each metaatom behaves as a Fabry–P\'erot–type resonator. Notably, the full \(2\pi\) phase swing arises from a resonance that spans the \(60\lambda_{\mathrm{Si}}\)-long modified region, and because the lateral pitch (300--500~nm) is \(\leq \lambda_{\mathrm{Si}}\), the higher diffraction orders are suppressed. Every metaatom therefore effectively behaves as a single-mode slab whose phase fully characterized by a single \(n_{\mathrm{eff}}\), eliminating spurious beams and ensuring that the collective phase response is dictated by sub-wavelength resonant coupling rather than simple geometric path length \cite{capasso_holey_lens}.

\section{Metalens Design for Continuous Nanofabrication}

\begin{figure}[H]
\centering
\includegraphics[scale=0.16]{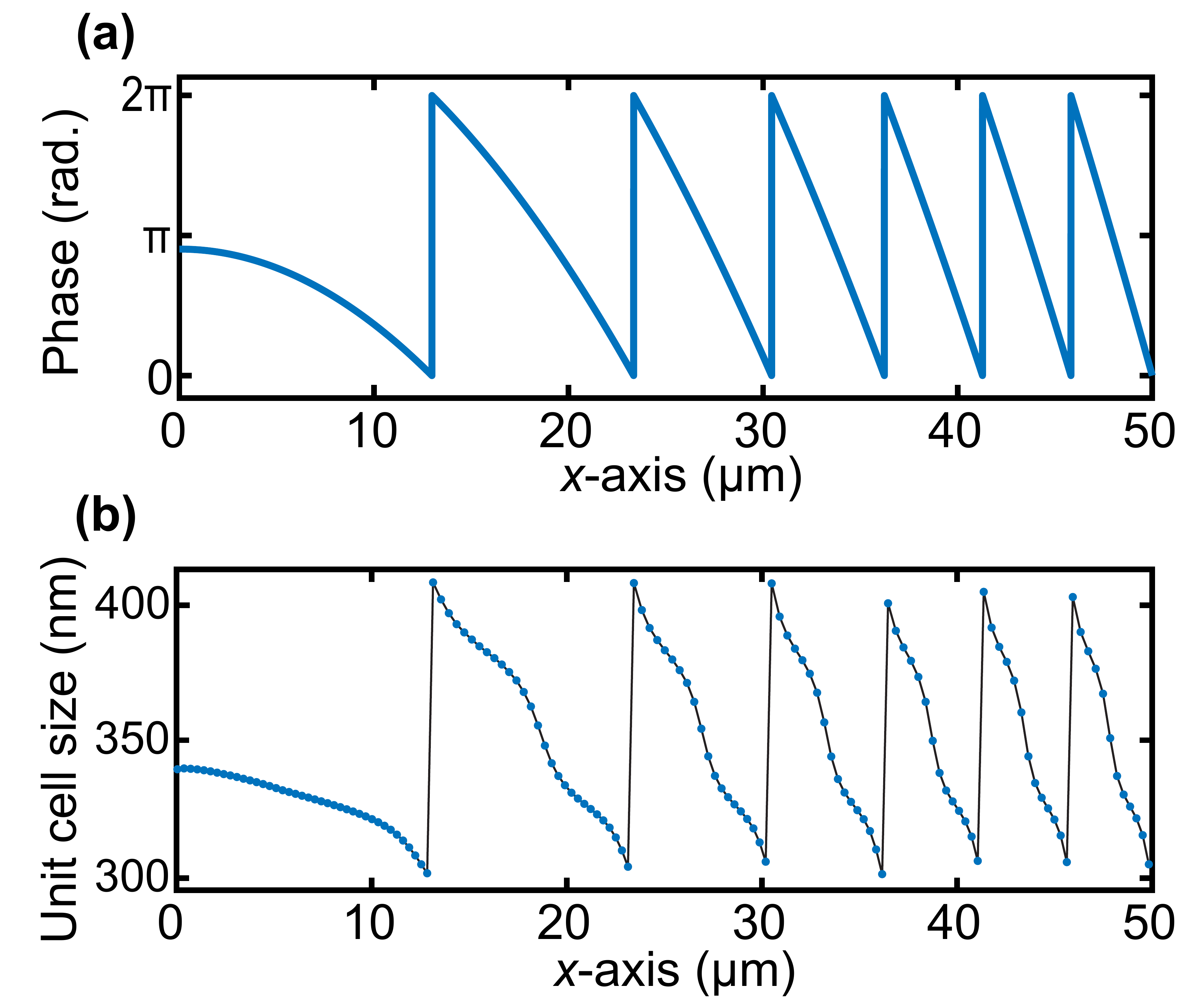}
\caption{ \textrm{\textbf{Phase mapping and discretization for the metalens.} (\textbf{a}) The target phase along the \textit{x} axis, found from the lens phase equation for $\varphi(r,\theta)$. (\textbf{b}) The phase is mapped to a unit cell distribution between 300 - 410 nm, computed with discretizing by \(\sim\)40 steps.}}
\label{fig:4}
\end{figure}

We design the metalens with the hyperbolic phase profile, \[\varphi(r,\theta) = 2\pi/\lambda \left( f - \sqrt{f^2 + r^2} \right)\] where $\lambda$ is the operating wavelength, $f$ the focal length, and ($r$, $\theta$) the polar coordinates in the lens plane. Fig. \ref{fig:4}a shows the wrapped phase along the \textit{x}-axis for a representative design with $f$ = 120 \textmu m and an aperture diameter of 100 \textmu m. By using the metaatom response given in Fig. \ref{fig:3}c, this phase profile is mapped to a unit cell distribution which in turn enables discretization (Fig. \ref{fig:4}b). This step converts the ideal phase map into a layout that can be written as a continuous, laser-inscribed array. Although the numerical discretization is performed with few nanometre-level pitch increments for accuracy, the resulting phase map may be quantized to the granularity of experiments.

An important design consideration is the polarization response of metaatoms. The procedure described in Fig. \ref{fig:4}b assumes an incident polarization angle of \(\theta=0^{\circ}\), with the metaatom width $\delta$l oriented perpendicular to the polarization. Since the mapping procedure is a function of $\theta$ (Supplementary Fig. S4), the preceding analysis can be repeated for the orthogonal case, \(\theta=90^{\circ}\). A continuous, polarisation-dependent phase profile is then obtained by fitting an elliptical interpolation curve in $\theta$ for each \(\Delta r_i\). Due to the slow variation of phase with orientation, this construction  captures the anisotropic phase response with high fidelity. The resulting loci in the ($r$, $\theta$) plane are only slightly elliptical and merge smoothly (Fig. \ref{fig:2}c), so the output phase of the metalens remains effectively circularly symmetric.

\begin{figure}[H]
\centering
\includegraphics[scale=0.17]{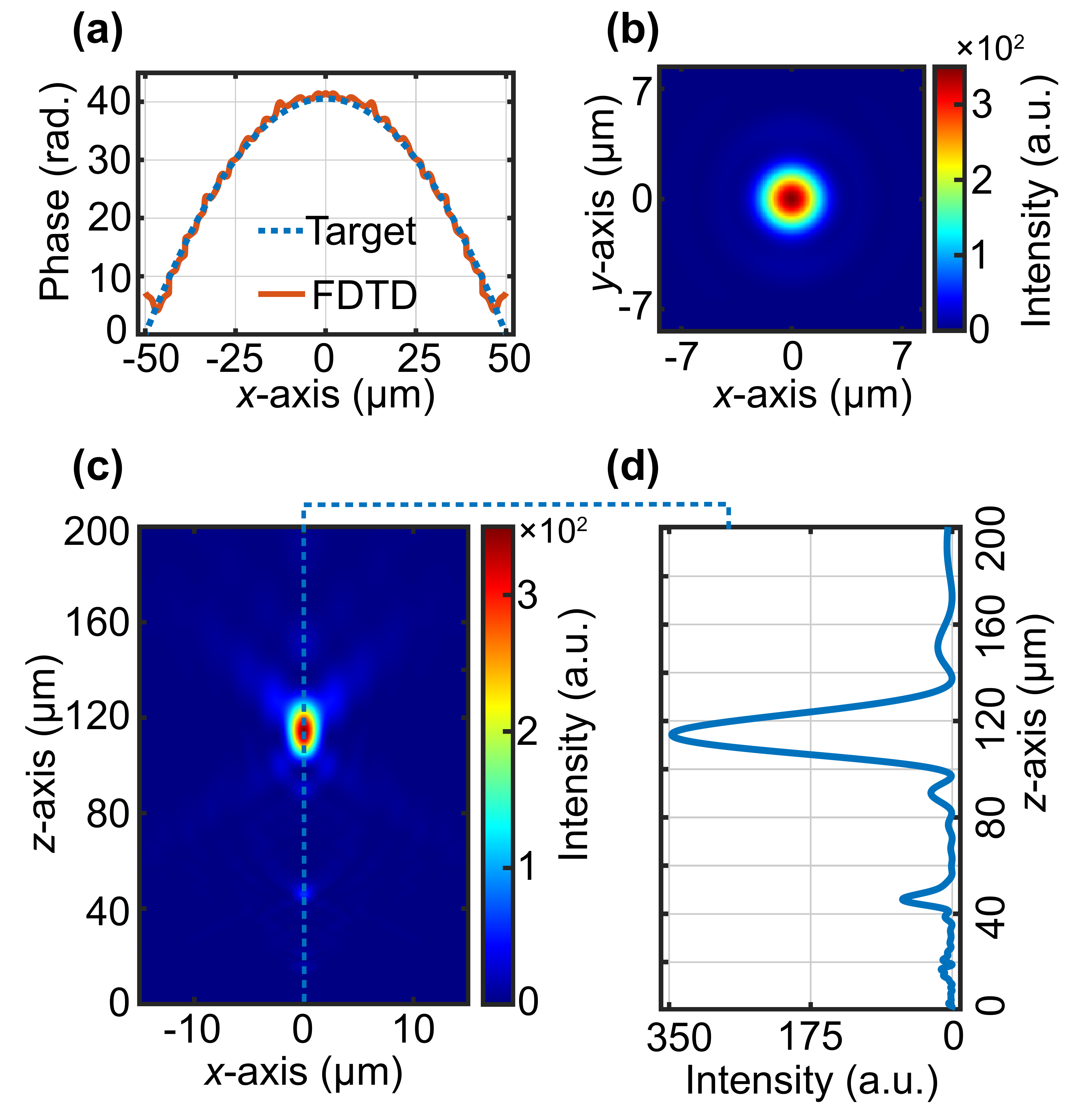}
\caption{\textrm{\textbf{In-chip metalens numerical characterization.} (\textbf{a}) Target phase front for a lens of $f$ = 120 \textmu m and aperture size 100 \textmu m, compared with the output (3D FDTD) of the designed element. (\textbf{b}) Intensity profile at the focal plane. (\textbf{c}) Propagation of the field in the $x-z$ plane. The metalens is positioned at $z=0$. (\textbf{d}) The $x=0$ cross-section of the field intensity after the metalens. Electric field is polarized along the \textit{x}-axis. Total simulation time of the metalens was near 1000 hours on a computer with Intel i7-10700F processor.}}
\label{fig:5}
\end{figure}

The complete metalens was simulated using a 3D FDTD solver. The resulting phase output closely matches the designed hyperbolic profile (Fig. \ref{fig:5}a), and the simulations confirm effective focusing behavior (Figs. \ref{fig:5}b--\ref{fig:5}d). We obtain a promising focusing efficiency of 70\% and a total transmission of 72\%. All transmission and focusing efficiency simulations assume ideal anti-reflection coatings at the input and output interfaces.

\section{Discussion}
To our knowledge, this study presents the first blueprint for subsurface metaoptics, achieving high efficiency through subwavelength phase control. Laser-written metaatoms were analyzed using a combination of FDTD simulations and a semi-analytical model to characterize their phase response. These results guided the design of a volumetric metalens architecture. The resonant configuration enables full 2$\pi$ phase modulation within a compact lateral footprint, supports fine phase quantization, and inherently suppresses the parasitic diffraction orders, features that distinguish it from non-resonant, path-length-based approaches. Experimental realizations will require precise and concurrent control over multiple fabrication parameters, including feature dimension, pitch, structure height, surface roughness, and refractive index modulation. While fine control over each parameter has been individually demonstrated in previous laser-writing studies\cite{Chambonneau2021_review,tokel2017chip,sabet2023laser}, device implementations will demand simultaneous, high-fidelity control within a single workflow. 

A critical direction for  advancing this platform is refractive index engineering. The proposed architecture assumes localized amorphization within silicon. Although full amorphization has not yet been achieved, recent ultrafast-laser writing studies report significant subsurface amorphization at the 100 nanometer scale \cite{Grojo2024PlasmaSeeds}. Moreover, structural analyses of laser-modified regions reveal the formation of nanoscale voids \cite{sabet2023laser}, suggesting the possibility of achieving an index contrast as high as 2.5. These developments open a broad design space for buried, high-performance metaoptics. Even the existing index contrast of $|\Delta n| \approx 10^{-2}$, as previously reported\cite{sabet2023laser}, may be sufficient. FDTD simulations in the 300$-$500 nm unit cell size range confirm that full $2\pi$ phase modulation can be achieved (Supplementary Fig. S5). In this regime, the reduced coupling strength is compensated by increasing the metaatom length up to 1200$\lambda_{\textrm{Si}}$, or approximately half a millimeter (Table S1). Although maintaining alignment over such aspect ratios is challenging, high-aspect-ratio structuring has already been demonstrated inside silicon \cite{tokel2017chip}. 

Device performance may be sensitive to fabrication imperfections, including variations in metaatom height, feature size, unit cell pitch, and refractive index. To assess robustness, we performed Monte Carlo simulations to evaluate phase sensitivity to these parameters (Table S2). The results indicate a fabrication-tolerant phase error of $\approx$ 0.5 radians. Minimizing scattering losses requires sidewall roughness below \(\lambda_{\text{Si}}/30\) \cite{capasso_holey_lens,Ma22}. Encouragingly, state-of-the-art laser-written modifications in silicon have achieved sidewall roughness near \(\lambda_{\text{Si}}/40\), which is promising for nanophotonic applications \cite{sabet2023laser}. Comparable tolerances apply to pitch and feature dimensions in current in-chip nano-fabrication \cite{sabet2023laser}. 

The current architecture encodes phase through unit cell size, discretized into forty levels (Fig. \ref{fig:4}b). Even a binary palette can produce a functional metalens, albeit with reduced efficiency \cite{binary_mopt}. Existing laser writing systems can generate five distinct levels, and further advances are anticipated to push this figure higher. Future efforts should aim to concurrently meet fabrication tolerances, sub$-$30 nm accuracy in height and pitch, sidewall roughness below \(\lambda_{\text{Si}}/30\), and unit cell quantization finer than today’s five‑level benchmark. These targets are within reach of current capabilities, and further improvements in beam shaping, feedback metrology, and post‑polish planarization may help close the remaining gap. 

Ultimately, the 3D Si-metaoptics paradigm holds the promise of multi-level monolithic integration for diverse functionalities. The maskless fabrication offers a direct, scalable and cost-effective path to buried subwavelength optics, devices that could operate alongside surface elements within standard CMOS flows \cite{sabet2023laser}. Given silicon’s transparency in the near- and mid-infrared, the platform can naturally extend to longer wavelengths. If fully 3D nanoscale architectures become accessible, one can envision spatial multiplexing of optical functions\cite{hu2024diffractive, abdollahramezani2020meta}. We further anticipate that this methodology will be transferable to other materials, particularly in light of recent advances in ultrafast laser nanostructuring of glasses and dielectrics \cite{Supstealth,ms_glass,Grojo_l_p_review,Yves:25}.

\section*{Data availability statement}

All data that support the findings of this study are included within the article (and any supplementary files).

\section*{Conflict of interest}
The authors have no conflict of interest.

\section*{Acknowledgment}

This study was partially supported by The Scientific and Technological Research Council of Turkey (TUBITAK) (Project no: 123M873).


\section*{References}
\bibliographystyle{iopart-num}
\bibliography{Metaoptics_abbreviated}  

\end{document}